\documentclass[preprint,twocolumn]{revtex4}
\usepackage{mathtools}
\usepackage{amsmath,bbm}
\usepackage{amssymb,bm}
\usepackage{array}
\usepackage{graphicx}
\usepackage{float}
\newcommand\nd{\noindent}
\begin{document}
\title{Temperature dependent formation-time approach for $\Upsilon$ suppression at energies available at the CERN Large Hadraon Collider}
\author{S. Ganesh\footnote{Corresponding author:\\Email: gans.phy@gmail.com}}
\author{M. Mishra}
\affiliation{Department of Physics, Birla Institute of Technology and Science, Pilani - 333031, Rajasthan, INDIA}
\vskip 3cm

\begin{abstract}
    We present here a comprehensive model to describe the bottomonium suppression data obtained from the CERN Large Hadron Collider (LHC) at center-of-mass energy of $\sqrt{s_{NN}}=2.76$ TeV. We employ a quasiparticle model (QPM) equation of state for the quark-gluon plasma (QGP) expanding under Bjorken's scaling law. The current model includes the modification of the formation time based on the temperature of the QGP, color screening during bottomonium production, gluon induced dissociation and collisional damping due to the imaginary part of the potential between the $b\bar b$ pair. We propose a method for determining the temperature-dependent formation time of bottomonia using the solution of the time-independent Schr\"{o}dinger equation and compare it with another approach based on time-dependent Schr\"{o}dinger wave equation simulation. We find that these two independent methods based on different axioms give similar results for the formation time. Cold nuclear matter effects and feed-down from higher resonance states of $\Upsilon$ have also been included in the present work. The suppression of the bottomonium states at mid rapidity is determined as a function of centrality. The results compare closely with the recent centrality-dependent suppression data at the energies available at the CERN LHC in the mid rapidity region. 

\vskip 0.5cm

{\nd \it Keywords } : Color screening, Gluonic dissociation, Collisional damping, Survival probability, CNM effects\\
{\nd \it PACS numbers } : 12.38.Mh, 12.38.Gc, 25.75.Nq, 24.10.Pa 

\end{abstract}

\maketitle
\section{Introduction}
   Chu and Matsui~\cite{Chu} employed the concept of color screening~\cite{mats} and presented a model to analyze the quarkonia suppression in quark-gluon plasma (QGP). Since then various experimental~\cite{PKSMCA,PKSRAR, PKSAAD, PKSCMS1,PKSBAB} and theoretical~\cite{PKSSPS1,PKSSPS2, PKSSPS3, PKSSPS4, PKSSPS5,PKSSPS6,PKSSPS7, yunpen,zhen,rishi,pksk} studies have been carried out on charmonium and bottomonium. Bottomonium suppression is considered as a more preferred probe in the Large Hadron Collider (LHC) experiments since its enhancement in the QGP medium is expected to be negligible even at the energies available at the LHC. 
The LHC data on bottomonium suppression~\cite{CMS2} at mid rapidity have been analyzed using a model based on color screening and gluonic dissociation along with collisional damping~\cite{gans}. However, in the above model, the formation time of bottomonia was taken independent of the temperature. This is not strictly valid.
At higher temperatures, the $Q\bar{Q}$ bonding is weakened due to Debye color screening, which leads to a decrease in the binding energy and hence an increase in the formation time. The formation time is a key ingredient for the color screening model. 
Its modification can significantly impact bottomonia suppression and consequently it is important to model the modification of formation time due to QGP temperature. 
The incorporation of the temperature effect on the formation time to model the quarkonia suppression is currently lacking in the literature.     
Here, we propose a method to estimate the formation time at different temperatures employing the solution of a time-independent Schr\"{o}dinger wave equation. We compare the outcome of this method with those obtained via the solution of a time dependent Schr\"{o}dinger wave equation. 
It is difficult to obtain the initial wavefunction for the time dependent Schr\"{o}dinger equation exactly. We heuristically arrive at the initial waveform. The heuristics and the associated uncertainties are described in detail in Sec. III (B).

It is expected that the gluonic dissociation and collisional damping play a prominent role only after the quarkonia is formed i.e., after the quarkonia formation time. However, due to increased formation time with temperature, the separation between $b$ and $\bar{b}$ is considerable even before $\Upsilon$ is fully formed. With a temperature-dependent formation time, the role of gluon-induced dissociation and collisional damping in bottomonia suppression before the bottomonium formation time may not be neglected. 
In view of the above facts, it becomes necessary to put forward a suppression model based on temperature-dependent formation time in order to analyze the experimental data. 

Furthermore, the cold nuclear matter (CNM) effects on bottomonia can contribute significantly at the energies available at the LHC.
Shadowing, nuclear absorption, and the Cronin effect are the main components of CNM effects. 
There have been studies on the initial state collisional and radiative energy loss in CNM, which can affect the nuclear modification factor, $R_{AA}$~\cite{cnmel0,cnmel1,cnmel2}. However, we do not include these initial-state energy-loss effects in the current work.
It has been argued that, as energy increases from the Super Proton Synchrotron (SPS) to the Large Hadron Collider (LHC) via the BNL Relativistic Heavy-ion Collider (RHIC) experiments, the absorption is expected to become less and less prominent~\cite{vogt}. The quark and antiquark in the $b\bar{b}$ pair immediately after being formed are very close to each other and hence the pair behaves almost as a color singlet and has negligible interaction with the nucleus, leading to almost no absorption. With higher energies, the $b\bar{b}$ pair size is expected to be even smaller when it crosses the nucleus leading to even smaller absorption. 
Because we describe $p_t$ integrated suppression data, the modeling of the $p_t$ broadening due to the Cronin effect is irrelevant.
This leaves shadowing as the principal CNM effect that needs to be incorporated.

There are various shadowing models available~\cite{EKS98, EKS99, nDSg, EPS08, EPS09} in the literature. For the purpose of modeling the shadowing effect, we explore two different approaches.
In the first approach, we employ the framework developed by Vogt~\cite{vogt} which is based on the shadowing parametrization first derived by Eskola et al.~\cite{EPS09}. From now onwards we refer to this shadowing parametrization as {\it EPS09}. 
In the second approach, which is a new approach proposed by us, we use the shadowing obtained by {\it EPS09} for the nuclei whose atomic mass is identical to or at least close to $\frac{N_{part}}{2}$. 
For instance, if $N_{part} = 32$, then each nuclei provides $16$ nucleons for the reaction. The effective shadowing would mainly be a function of these $16$ nucleons. Interestingly, we find that the two approaches yield similar results. The shadowing effect, being related to the parton distribution function in the heavy nuclei, is expected to be an initial-state effect and should have the same or similar effect independent of the final state of heavy quarkonium ($1S$ or $2S$ etc). We use the same shadowing effect in all the final bottomonium states. 

 In the current work, we describe a comprehensive model by incorporating the effect of temperature on formation time, color screening, gluon-induced dissociation, and collisional damping along with CNM effects. All these effects are modeled analytically. 
The results of the model are compared with the recent centrality dependent CMS data~\cite{CMS2} at mid rapidity obtained from the LHC experiments. The organization of the rest of the article is as follows. Section II, briefly describes the color screening, gluonic dissociation, and collisional damping that have been used in this work. Section III describes how the temperature dependent formation time has been modeled. CNM effects are also described in this section. Section IV gives the results and the comparison with the CMS data. Finally, we conclude our work in Sec. V.

\section{Color screening, Gluon dissociation and Collisional damping}
\subsection{Color screening}
The color screening model used in the present work is based on the pressure profile~\cite{Madhu1} in the transverse plane and the cooling law for pressure based on a quasiparticle model (QPM) equation of state (EOS)~\cite{Madhu2} for the QGP. It is described in detail in Refs. ~\cite{gans,Madhu2}. The cooling law for pressure is given by
\begin{equation}
	p(\tau,r) = A + \frac{B}{\tau^q} + \frac{C}{\tau} + \frac{D}{\tau^{c_s^2}},
\end{equation}
where A = -$c_1$, B = $c_2c_s^2$, C = $\frac{4\eta q}{3(c_s^2 - 1)}$ and D = $c_3$. The constants $c_1$, $c_2$ and $c_3$ are given by
\begin{itemize}
\item $c_1 = - c_2\tau'^{-q} - \frac{4\eta}{3c_s^2\tau'}$,
\item $c_2 = \frac{\epsilon_0 - \frac{4\eta}{3c_s^2}\left( \frac{1}{\tau_0} - \frac{1}{\tau'} \right ) }{\tau_0^{-q} - \tau'^{-q}}$,
\item $c_3 = \left (p_0 + c_1 \right) \tau_0^{c_s^2} - c_2c_s^2 \tau_0^{-1} - \frac{4\eta}{3}\left ( \frac{q}{c_s^2 - 1} \right ) \tau_0^{\left ( c_s^2 - 1\right )}$.
\end{itemize}
The above constants are determined by using different boundary conditions on pressure and energy density described in Refs. ~\cite{gans,Madhu2}. 

Writing Eq. (1) at initial time $\tau = \tau_i$ and screening time $\tau = \tau_s$ and combining it with the pressure profile~\cite{Madhu2}, we get the following two equations:

\begin{equation}
	p(\tau_i,r) = A + \frac{B}{\tau_i^q} + \frac{C}{\tau_i} + \frac{D}{\tau_i^{c_s^2}} = p(\tau_i,0)\,h(r),
\end{equation}
\begin{equation}
	p(\tau_s,r) = A + \frac{B}{\tau_s^q} + \frac{C}{\tau_s} + \frac{D}{\tau_s^{c_s^2}} = p_{QGP},
\end{equation}
where $p_{QGP}$ is the pressure of QGP inside the screening region required to dissociate a particular $\Upsilon$ state and it is determined by a QPM EOS for QGP medium. The above equations are solved numerically and we equate the screening time $\tau_s$ to the dilated formation time of bottomonia $t_F = \gamma \tau_F(T)$ to determine the radius of the screening region $r_s$~\cite{gans,Madhu2}.
The $\tau_F(T)$ is now dependent on the QGP temperature $T$. $\gamma = \frac{E_T}{M_{\Upsilon}}$ is the Lorentz factor corresponding to the transverse energy $E_T$ and the bottomonium mass $M_{\Upsilon}$. 
The expression for survival probability due to color screening can be obtained as: 
\begin{eqnarray}
	S_c(p_T,N_{part}) = \frac{2(\alpha + 1)}{\pi R_T^2} \int_0^{R_T} dr\, r \,\phi_{max}(r)\\\nonumber 
\left \{ 1 - \frac{r^2}{R_T^2} \right \}^\alpha, 
\end{eqnarray}
where $\alpha=0.5$, and $R_T$ and $\phi_{max}$ (which is a function of $p_t$ and $r_s$) are defined in Refs. \cite{gans,Madhu2}.
The above expression for survival probability is integrated over the range of $p_t$ allowed by the CMS experiment~\cite{CMS2} to obtain the $p_t$ integrated survival probability. 
\subsection{Collisional damping}

The singlet potential used in this work is given by~\cite{Wolschin}
\begin{equation} 
\begin{split}
	V(r,m_D) = \frac{\sigma}{m_D}(1 - e^{-m_D\,r}) - \\ 
\alpha_{eff} \left ( m_D + \frac{e^{-m_D\,r}}{r} \right ) - \\
i\alpha_{eff} T \int_0^\infty \frac{2\,z\,dz}{(1+z^2)^2} \left ( 1 - \frac{\sin(m_D\,r\,z)}{m_D\,r\,z} \right ),
\end{split}
\end{equation} 
where

\begin{itemize}
\item $m_D$ is the Debye mass given by 
\begin{math}
	m_D = T\sqrt{4\pi \alpha_s^T \left ( \frac{N_c}{3} + \frac{N_f}{6} \right ) }.
\end{math}
\item $\alpha_{eff} = \frac{4\alpha}{3} = 0.63$, $N_f = 3$ = number of flavors, $\alpha_s^T = 0.47$, and $\sigma= 0.192$ GeV$^2$.
\end{itemize}

The collisional damping dissociation time constant is given by
$\Gamma_{damp} = \int[\psi^\dagger \left [ Im(V)\right ] \psi]$\,dr~\cite{gans}, with $\psi$ being the bottomonium wave function.

We solve the Schr\"{o}dinger equation to get the radial wave function for the $1S$, $2S$ and $1P$ states.

\subsection{Gluonic dissociation}
 We model the gluonic dissociation cross section as~\cite{Wolschin}:
\begin{equation}
\begin{split}
\sigma_{diss,nl}(E_g) = \frac{\pi^2\alpha_s^u E_g}{N_c^2} \sqrt{\frac{m}{E_g + E_{nl}}} \\
\left ( \frac{l|J_{nl}^{q,l-1}|^2 + (l+1)|J_{nl}^{q,l+1}|^2}{2l+1} \right), 
\end{split}
\end{equation}
with $\alpha_s^u = 0.59$,
and $J_{nl}^{ql'}$ can be expressed using singlet and octet wave functions as:
\begin{equation}
J_{nl}^{ql'} = \int_0^\infty dr\,r g^*_{nl}(r)h_{ql'}(r).
\end{equation}
The octet wave function $h_{ql'}$ is obtained by solving the Schr\"{o}dinger equation with potential, $\alpha_{eff}/(8\,r)$~\cite{Wolschin,Octet,gans}.
The cross section is then averaged over a Bose-Einstein distribution function of gluons at temperature $T$ as:
\begin{equation}
\Gamma_{diss,nl} = \frac{g_d}{2\pi^2} \int_0^\infty \frac{dp_g\,p_g^2 \sigma_{diss,nl}(E_g)}{e^{E_g/T} - 1},
\end{equation}
with $g_d$ = $16$ for gluons.

The net dissociation constant is given by 
\begin{equation}
    \Gamma_{total} = \Gamma_{damp} + \Gamma_{diss}.
\end{equation}
At the very initial time, the $b\bar{b}$ pair would be very close to each other and hence would behave almost as a color singlet. During this time, the gluon-induced dissociation and the dissociation due to collisional damping may be assumed to be negligible. Consequently, we integrate the dissociation due to the above two processes from a non-zero initial time, $t_0$.  
The survival probability due to gluonic dissociation along with collisional damping is then given by
\begin{equation}
S_g=\int_{t_0}^{\infty}exp(-\Gamma_{total})\,dt.
\end{equation} 
We take the value of $t_0$ to be $0.5$ fm.

\subsection{Net survival probability}
From Ref.~\cite{gans}, the net survival probability is obtained by combining $S_c$ and $S_g$
\begin{center}
\begin{math}
S' = S_c\,S_g.
\end{math}
\end{center}
The expressions for survival probability after incorporating feed-down corrections are expressed as: 

\begin{eqnarray}
S_{1S} = 0.6489\,S'_{1S} + 0.1363\,S'_{1P} \nonumber \\
+ 0.1733 \, S'_{2S} + 0.0416 \,S'_{2P},\nonumber 
\end{eqnarray}
\begin{eqnarray}
S_{1P} = 0.8450\, S'_{1P} + 0.1508 \, S'_{2S} + \nonumber \\
0.0041 \, S'_{2P}, \nonumber
\end{eqnarray}
\begin{eqnarray}
S_{2S} = 0.8780 \, S'_{2S} + 0.1220 \, S'_{2P},\nonumber \\
\end{eqnarray}
where $S'_{nl}$ is the survival probability of the $|nl\rangle$ quarkonium states before feed-down is considered, while $S_{nl}$ is the survival probability of the $|nl\rangle$ state after feed-down. 

\section{Temperature dependent formation time and CNM effects}
\subsection{Temperature dependent formation time}
We model the temperature as a function of $N_{part}$ as~\cite{gans}: 
\begin{equation}
T(t_{qgp}) = T_c \left ( \frac{N_{part}(bin)}{N_{part}(bin_0)}\right )^{1/3} \left ( \frac{t_{QGP}}{t_{qgp}} \right )^{1/3}
\end{equation} 
In the above equation, $T_c = 0.170$ GeV, $t_{QGP}$ is the total QGP lifetime taken as $5$ fm, and $t_{qgp}$ is the current time of the QGP.
The temperature is inversely proportional to the cube root of proper time. The cube root can be seen from the QPM as a limiting case when the Reynolds number R $\gg$ 1~\cite{gans}.
To model the temperature dependence of the formation time we use the real part of the singlet potential given in Eq. (5),
namely 
\begin{equation} 
\begin{split}
	V(r,m_D) = \frac{\sigma}{m_D}(1 - e^{-m_D\,r}) - \\ 
\alpha_{eff} \left ( m_D + \frac{e^{-m_D\,r}}{r} \right ).
\end{split}
\end{equation} 
Solving the Schr\"{o}dinger equation 
$-\frac{1}{2\mu}\frac{\partial^2\psi}{\partial^2r}+V(r,m_D)\psi + \frac{l(l+1)}{2\mu r^2}\psi=E_T(n,l)\psi$, 
gives the energy eigenvalues $E_T(n,l)$ ($n$ = principal quantum number, $l$ = azimuthal quantum number) at temperature $T$ for the bottom quark antiquark system with reduced mass $\mu$. 
We calculate the binding energy as:
\begin{equation}
E_{bind}(T) =  E_T(n,l) - V(r= \infty,m_D(T)).
\end{equation}
The formation time of $\Upsilon(1S)$, $\Upsilon(1P)$ and $\Upsilon(2S)$ at $0$ MeV (in vacuum) are taken as $0.76$, $2.6$ and $1.9$ fm, respectively~\cite{formt0, formt1, formt2, formt3}. The binding energy values in a vacuum are taken as $1.1$, $0.67$ and $0.54$ GeV, respectively~\cite{binde1,binde2}. 
One can note that the given formation times are greater than the inverse of the binding energies in vacuum. In general, the formation time is taken to be of the form $\frac{K}{E_{bind}(T)}$, with $K \ge 1$.
One can then determine the formation time at a temperature $T_1$ as: 
\begin{equation}
\tau_f(T_1) = \tau_f(T_0)\frac{E_{bind}(T_0)}{E_{bind}(T_1)}.
\end{equation} 
The above equation indicates that if the formation time in vacuum (i.e., at temperature $T_0 = 0$ MeV) is known, then the formation time at any other temperature $T_1$ can be calculated.
We solve a time independent Schr\"{o}dinger equation and determine the $\Upsilon$ wavefunction $\psi_T(r)$ for various temperatures $T$.
Equation (12) shows that the temperature of the QGP decreases as the cube root of proper time. The question then arises as to what temperature needs to be used to determine the formation time at a particular centrality. 
  We suggest to use the temperature $T$ such that $\tau_f[T(t_{qgp})] = t_{qgp}$, where $t_{qgp}$ is the time for which the QGP has existed. The motivation behind this is as follows: When the $b\bar{b}$ pair is formed, it will keep on expanding until the stable state of $\Upsilon$ is formed. Let at QGP time $t_0$, the temperature of the QGP be $T(t_0)$. If $t_0 < \tau_f[T(t_0)]$ (given by Eq. (15)), then the $\Upsilon$ has not yet reached its stable state and will continue to expand. The desired formation time would then be greater than $t_0$. As a consequence of the above, a point in time will be reached when $t_0 = \tau_f[T(t_0)]$. At this point the $b\bar{b}$ will be at equilibrium and the $\Upsilon$ will just get formed. Beyond this time the temperature will decrease and the $\Upsilon$ wavefunction may shrink due to string tension, but will continue to remain stable unless dissociated by the sea of gluons present in the QGP or by collisions with other particles in the QGP.
The condition $\tau_f[T(t_{qgp})] = t_{qgp}$ may not be valid for $Q\bar{Q}$ created much after the formation of the QGP. As a corollary, the ability of this temperature dependent formation time model to explain CMS data gives an indication that the $\Upsilon$ enhancement may be negligible at the energies available at the LHC.
We solve the equation $\tau_f[T(t_{qgp})] = t_{qgp}$ by finding the intersection of the two curves $t_{qgp}$ and $\tau_f[T(t_{qgp})]$. Figure 1 shows the two curves for $\Upsilon(1S)$ corresponding to the most central bin. The point of intersection gives the formation time as $2.35$ fm for $\Upsilon(1S)$ for the most central bin. 
\begin{figure}[h!]
\includegraphics[width = 80mm,height = 80mm]{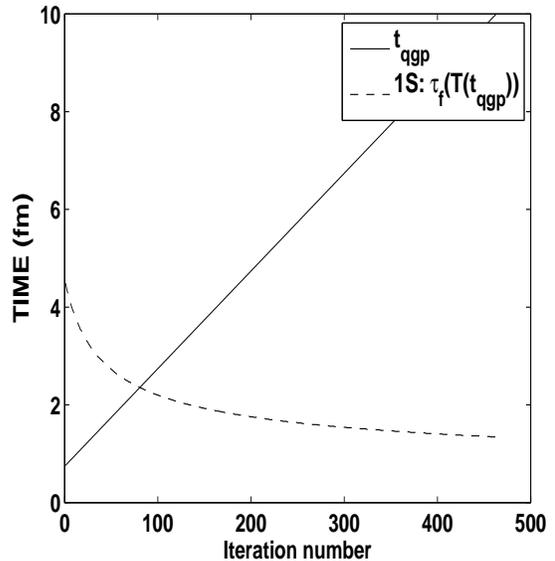}
\caption{Formation time given by the intersection of the two curves $t_{qgp}$ and $\tau_f[T(t_{qgp})]$ for $\Upsilon(1S)$ for the most central bin.}
\label{fig:formation_time}
\end{figure}
\subsection{Formation time based on a time-dependent Schr\"{o}dinger wave equation}
 We now compare the above method of determining the formation time with another one based on the solution of a time-dependent Schr\"{o}dinger wave equation: 
\begin{math}
	\psi_T(r,t_N) = \left [ \Pi^{N-1}_{i=0} e^{iH(t_i)(t_{i+1} - t_i)}\right ] \psi_T(r,t_0),
\end{math}
where $H(t_i)$ is the Hamiltonian with the potential term given in Eq. (13), along with the kinetic energy and $\frac{l\,(l+1)}{2\mu\,r^2}$ terms. The potential is evaluated at temperature $T(t_i)$ of the QGP at time $t_i$.
To determine the formation time from simulation, it is required to use the following:  
\begin{itemize}
\item a criteria to define the condition that the $\Upsilon$ has formed.
\item An initial condition for the simulation.
\end{itemize}
For the purpose of determining the formation time from the simulation results, we use the following criteria: 
If the $\Upsilon$ wavefunction $\psi(r,t)$ is normalized such that,  
\begin{math}
	\int \psi_T^{\dagger}(r,t)  \psi_T(r,t) dr~=~1,
\end{math}
then the mean distance $\langle r(t) \rangle$ between the quark and anti-quark can be written as:
\begin{math}
	\langle r(t) \rangle = \int \psi_T^{\dagger}(r,t) r \psi_T(r,t) dr.
\end{math}
~(Fig. 2)
We define the formation time as the earliest time at which the $\Upsilon$ becomes stable. We say that the $\Upsilon$ has become stable when the wavefunction ceases to expand, i.e., when
\begin{math}
	\frac{d\langle r(t)\rangle}{dt}~=~0.
\end{math}

Because the effect of temperature can only widen the wavefunction and thus increase the formation time (as compared to formation time in vacuum), 
the simulation for a particular state of $\Upsilon$ is started at a time equal to the formation time $t_{f0}$ for that state in vacuum. 
At any point of time, the $\Upsilon$ wavefunction influenced by the QGP medium would be expected to have much larger separation between the quark and antiquark, as compared to the quark-antiquark separation within the bottomonium wavefunction in vacuum which has evolved for the same amount of time. This is realized by the non zero value of the temperature $T_{init}$, at which the initial wavefunction is determined. 
In fact, for this reason, there is a need to apply a lower bound on the value of $T_{init}$ to a value higher than $0$ MeV. 
In view of the above arguments, we heuristically fix the lowest value of $T_{init}$ to a nominal value of $30$ MeV.
For the initial wavefunction at the start of the simulation, we propose to use the time independent wavefunction $\psi_T(r)$ at the temperature $T_{init}$, subject to the lower bound of $30$ MeV, for which  the criteria $\frac{d<r(t)>}{dt}~=~0$ exists. Below the temperature $T_{init}$, the wavefunction keeps on expanding. 

For the $\Upsilon(1S)$ state, we find that the $\frac{d<r(t)>}{dt}=0$ criteria exists when we start with $\psi_T(r,t_0)$ at any small value of temperature $T$, including $T = 0$ MeV. Based on the above arguments, we choose $T_{init} = 30$ MeV. However, for $\Upsilon(2S)$ and $\Upsilon(1P)$, the value of temperature varies from $T_{init}~=~160$ MeV to $T_{init}~=~30$ MeV depending upon the centrality. 


\begin{figure}[h!]
\includegraphics[width = 80mm,height = 80mm]{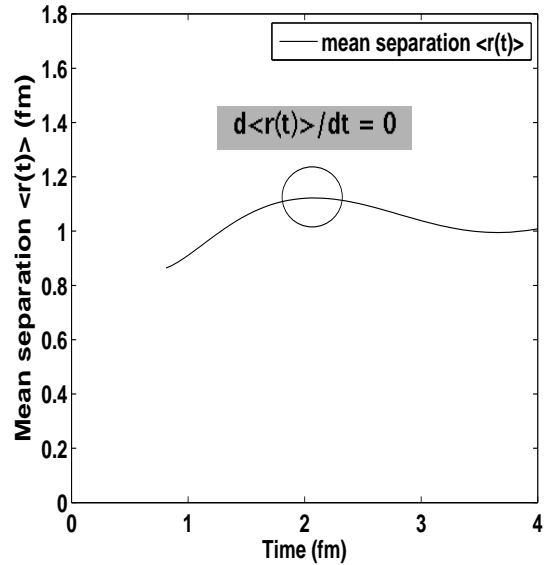}
\caption{Variation of mean separation between the $b\bar{b}$ pair with time for $\Upsilon(1S)$ for the most central bin.} 
\label{fig:mean_r_1S}
\end{figure}

With the above framework for the time dependent Schr\"{o}dinger equation, the simulation results for the three states of $\Upsilon(1S)$, $\Upsilon(2S)$ and $\Upsilon(1P)$ are shown in Figs. 3, 4 and 5, respectively.

\begin{figure}[h!]
\includegraphics[width = 80mm,height = 80mm]{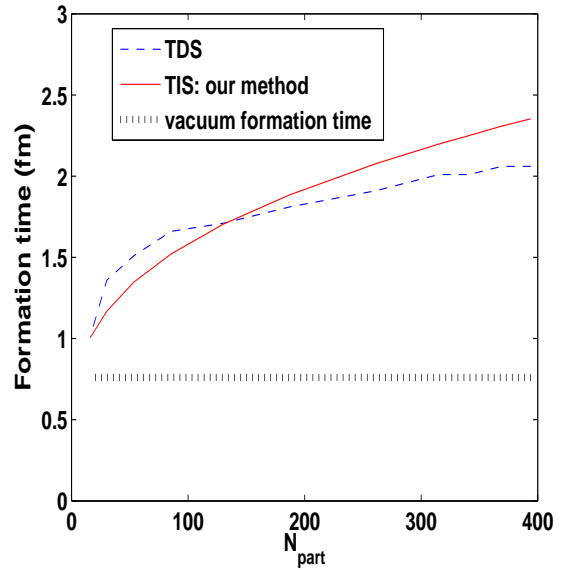}
\caption{Comparison of the $\Upsilon(1S)$ formation time obtained using the time-dependent Schr\"{o}dinger equation (TDS), our method based on the time-independent Schr\"{o}dinger equation (TIS) and the vacuum formation time.}
\label{fig:1S Formation time }
\end{figure}

\begin{figure}[h!]
\includegraphics[width = 80mm,height = 80mm]{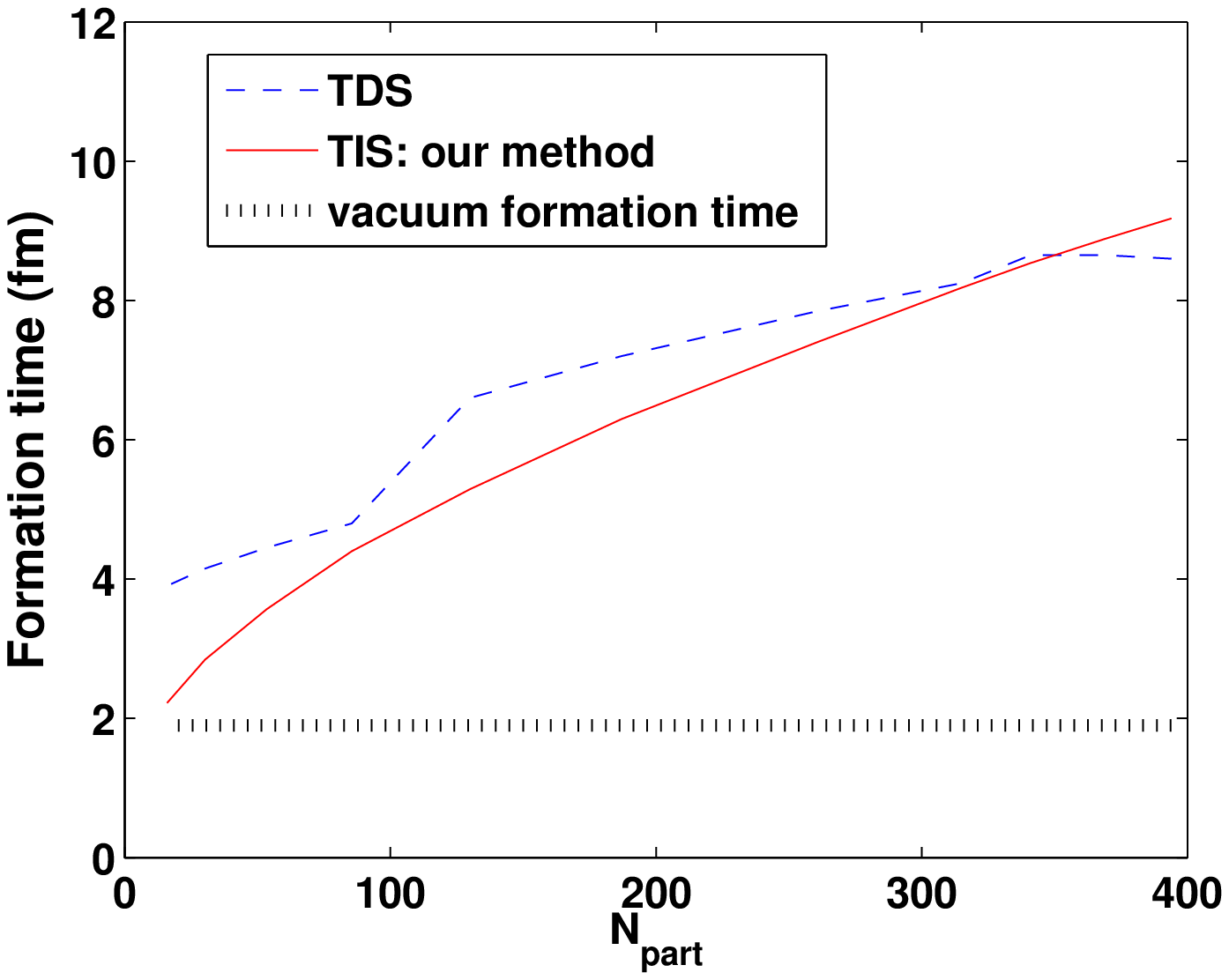}
\caption{Comparison of the $\Upsilon(2S)$ formation time obtained using the time-dependent Schr\"{o}dinger equation (TDS), our method based on the time-independent Schr\"{o}dinger equation (TIS), and the vacuum formation time.}
\label{fig:2S Formation time}
\end{figure}

\begin{figure}[h!]
\includegraphics[width = 80mm,height = 80mm]{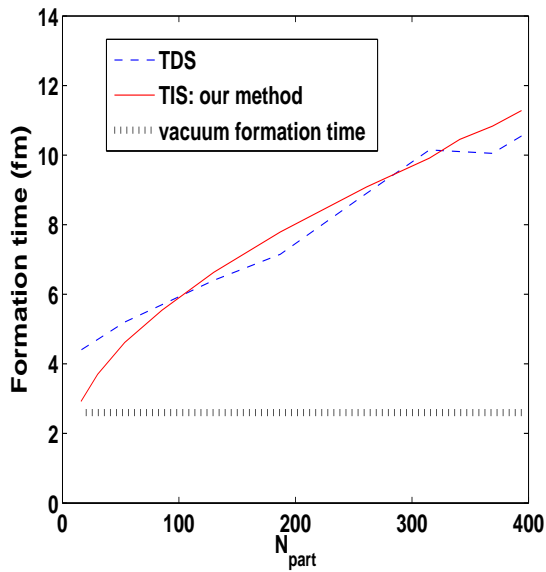}
\caption{Comparison of the $\Upsilon(1P)$ formation time obtained using the time dependent Schr\"{o}dinger equation (TDS), our method based on the time-independent Schr\"{o}dinger equation (TIS), and the vacuum formation time.}
\label{fig:1P Formation time }
\end{figure}

We find that the method based on the time-dependent Schr\"{o}dinger equation follows the trend of the formation time as a function of centrality as determined by our proposed method, i.e., by solving the time-independent Schr\"{o}dinger equation.
Differences in the two curves could possibly be due to the uncertainty in the initial conditions. For central collisions, the time interval for which the time-dependent Schr\"{o}dinger equation is simulated is large (since the formation time is large), and errors in the initial conditions probably have a lesser impact. But for peripheral collisions where the time interval of simulation is much smaller, the impact of the initial conditions is much larger. 
In the results section, we describe the $\Upsilon$ suppression by using both the sets of formation time and we show that the difference is almost negligible. At the same time, we also show that the suppression obtained by ignoring the effect of temperature $T$ on the formation time and using the $T$-dependent formation time is significantly different.
\subsection{CNM effects}
As mentioned in the introduction, nuclear absorption is expected to be quite small and the Cronin effect irrelevant, the main CNM effect is therefore shadowing. Thus, only the shadowing effect has been used to determine the contribution of suppression arising due to the CNM effect.
 Shadowing is an initial-state effect and hence it is expected to be similar for all the bottomonium states~\cite{vogt}. 
  We now describe the two approaches that we have used to model shadowing in the current work.

\subsubsection{The Vogt approach}
Vogt~\cite{vogt} has computed the shadowing effect at a center-of-mass energy of $\sqrt{s_{NN}} = 5.5$ TeV. 
We employ the same formulation to compute the shadowing effect for various centrality bins at $\sqrt{s_{NN}} = 2.76$ TeV. We use the {\it EPS09}~\cite{EPS09} parametrization to obtain the shadowing $S^i(A,x,\mu)$ for nucleus with atomic mass number $A$, momentum fraction $x$, and scale $\mu$. The superscripts $i$ and $j$ refer to the incoming and target nuclei, respectively.

The spatial variation of shadowing $S^i_{\rho}(A,x,\mu,\vec{r},z)$, is taken to be a function of the shadowing $S^i(A,x,\mu)$ and the nucleon density $\rho_A(\vec{r},z)$ as:
\begin{eqnarray}
S^i_{\rho}(A,x,\mu,\vec{r},z) = 1 + N_{\rho}(S^i(A,x,\mu)-1)\\ \nonumber 
\frac{\int dz\rho_A(\vec{r},z)}{\int dz\rho_A(0,z)},
\end{eqnarray}
where $N_{\rho}$ is determined by the following normalization condition:
\begin{equation}
\frac{1}{A} \int d^2rdz\rho_A(s)S^i_{\rho}(A,x,\mu,\vec{r},z) = S^i(A,x,\mu),
\end{equation}
with atomic mass number $A = 208$ for Pb and $s = \sqrt{r^2 + z^2}$.
The nuclear density $\rho(s)$ has been taken to be the Woods Saxon distribution:
\begin{math}
\rho_A(s) = \rho_0 \frac{1 + \omega(s/R_A))^2}{1 +  exp[(s- R_A)/d]}.
\end{math}
The values of $\rho_0$, $R_A$, $d$ and $\omega$ have been taken from Ref.~\cite{woods} for Pb.

The suppression factor is defined as the ratio:
\begin{equation}
R_{AB}(N_{part};b) = \frac{d\sigma_{AB}/dy}{T_{AB}(b)d\sigma_{pp}/dy},
\end{equation}
where $b$ is the impact parameter and $T_{AB}$ is the nuclear overlap function given by 
\begin{equation}
	T_{AB}(b) = \int d^2s\,dz_1\,dz_2\rho_A(\vec{s},z_1)\rho_B(|\vec{b}-\vec{s}|,z_2).
\end{equation}
In this particular case, both $A$ and $B$ stand for atomic mass numbers of Pb.

From Ref.~\cite{newvogt}, the color evaporation model gives
\begin{equation}
\begin{split}
\sigma_{AB} = \int dz_1 \,dz_2\, d^2r\, dx_1\, dx_2\,[f^i_g(A,x_1,\mu,r,z_1)\\
f^j_g(B,x_2,\mu,b-r,z_2) \sigma_{gg\_QQ}(x_1, x_2, \mu)],
\end{split}
\end{equation}
\begin{equation}
\begin{split}
	\sigma_{pp} = \int dx_1\,dx_2\,[f_g(p,x_1,\mu)\\
	f_g(p,x_2,\mu) \sigma_{gg\_QQ}(x_1, x_2, \mu)].
\end{split}
\end{equation}
The above formalism excludes the explicit modeling of spin and color of the initial and final states.

In the above expressions, $x_1$ and $x_2$ are the momentum fraction of the gluons in the two Pb nuclei at $\sqrt{s_{NN}} = 2.76$ TeV and are related to the rapidity $y$ as:
\begin{math}
	x_1 = \frac{M_t}{\sqrt{s_{NN}}}e^{y}
\end{math}
and 
\begin{math}
	x_2 = \frac{M_t}{\sqrt{s_{NN}}}e^{-y},
\end{math}
with $M_t = \sqrt{M_{\Upsilon}^2 + \langle p_t \rangle^2}$, where $M_{\Upsilon}$ is the bottomonium mass and $\langle p_t \rangle$ is the mean transverse momentum of the bottomonium.
\\
The function $f_g(A,x,\mu,r,z)$, is determined from the gluon distribution function for proton $f_g(p,x,\mu)$, by using the following relation:\\
\begin{itemize}
\item
\begin{math}
f^i_g(A,x_1,\mu,r,z) = \rho_A(s)S^i(A,x_1,\mu,r,z)f_g(p,x_1,\mu)
\end{math}
\item
\begin{math}
f^j_g(B,x_2,\mu,b-r,z) = \rho_B(s)S^j(B,x_2,\mu,b-r,z)f_g(p,x_2,\mu)
\end{math}
\end{itemize}
By using the above expressions, the value of $\frac{d\sigma_{AB}}{d y}$ and $\frac{d\sigma_{pp}}{d y}$ can be computed numerically and, finally, $R_{AB}$ can be obtained.


The value of the gluon distribution function $f_g(p,x,\mu)$ in a proton (indicated by label $p$) has been estimated by using {\it CTEQ6}~\cite{CTEQ6}. 
\subsubsection{Our approach}
Repeating the example mentioned in the Introduction, if $N_{part} = 32$, then each nuclei provides $16$ nucleons for the reaction. The effective shadowing is taken to be mainly a function of these $16$ nucleons. In general, 
\begin{equation}
	S^i(A=208,x,\mu,b) = S(A = \frac{N_{part}}{2},x,\mu),
\end{equation}
where $b$ is the impact parameter corresponding to the given value of $N_{part}$ and $S(A = \frac{N_{part}}{2},x,\mu)$ is the value obtained from {\it EPS09} for nuclei with atomic mass $N_{part}/2$.
We use this procedure to determine the shadowing $S^i(A,x,\mu,b)$ and $S^j(A,x,\mu,b)$ for the various centrality bins. 

Once $S^i(A,x,\mu,b)$ and $S^j(A,x,\mu,b)$ are determined, we determine $\sigma_{AB}$ as
\begin{equation}
\begin{split}
\sigma_{AB} = \int  dx_1\, dx_2\,f^i_g(A,x_1,\mu,b)\\
\times f^j_g(B,x_2,\mu,b) \sigma_{gg\_QQ}(x_1, x_2, \mu),
\end{split}
\end{equation}
where\\
\begin{math}
f^i_g(A,x_1,\mu,b) = S^i(A,x_1,\mu,b)f_g(p,x_1,\mu)
\end{math}
\\and\\
\begin{math}
f^j_g(B,x_2,\mu,b) = S^j(B,x_2,\mu,b)f_g(p,x_2,\mu).
\end{math}
\\
\\
We then use Eqs. (18) and (19) to determine $R_{AB}$.
The shadowing data are not available for all the atoms, which leads to the possibility that, for a particular centrality bin, the corresponding shadowing data for atoms with atomic mass $\approx$ $N_{part}/2$ would not be available. In such cases we use linear interpolation of the available shadowing data of the atoms with the closest value of atomic mass. 
We justify the use of linear interpolation by the fact that the shadowing effect is almost a monotonic smooth function of atomic mass (see Fig. 6).
\begin{figure}[h!]
\includegraphics[width = 80mm,height = 80mm]{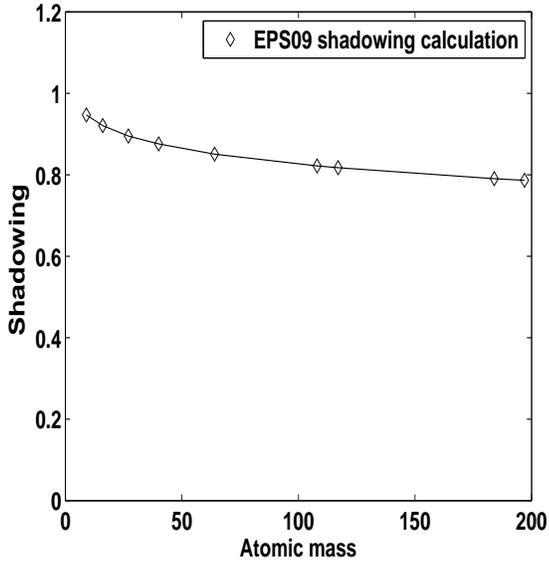}
\caption{Monotonic and smooth nature of shadowing versus atomic mass. The shadowing values shown here are for momentum fraction $= 0.000312$.}
\label{fig:monotonic}
\end{figure}

The atomic mass (more appropriately, we call it effective atomic mass) used for each value of $N_{part}$ is given in Table I. 	
\begin{table}
\caption{Values of the effective atomic mass and corresponding $N_{part}$.}
\begin{tabular}{|c|c|c|}
\hline\hline
Centrality & $N_{part}$ & Effective atomic mass \\
bin &       &\\
\hline
1  & 394 & 197  \\
\hline
2  & 369 & 184  \\
\hline
3  & 341 & linear interpolation of 108 and 184  \\
\hline
4  & 315 & linear interpolation of 108 and 184  \\
\hline
5  & 261 & linear interpolation of 108 and 184  \\
\hline
6  & 187 & linear interpolation of 64 and 108  \\
\hline
7  & 130 & 64 \\
\hline
8  & 85.5 & 40 \\
\hline
9  & 53 & 27 \\
\hline
10  & 30.3 & 16 \\
\hline
11  & 16 & 9 \\
\hline\hline
\end{tabular}
\end{table}

The values of the parameters used in our simulations are given in Table II.
\begin{table}
\caption{Values of the input data used in our simulation~\cite{Wolschin}.}
\begin{tabular}{|l|l|l|l|l|}
\hline\hline
$\Upsilon$ property & $\Upsilon(1S)$ & $\Upsilon(2S)$ & $\Upsilon(1P)$  & $\Upsilon(2P)$ \\
\hline
Mass (GeV) & 9.46 & 10.02 & 9.99 & 10.26 \\
\hline
$\tau_F$ (fm) & 0.76 & 1.9 & 2.6 & - \\
\hline
$T_{diss}$ (MeV) & 668 & 217 & 206 & 185$^*$\\
\hline\hline
\end{tabular}
\end{table}

\begin{figure}[h!]
\includegraphics[width = 80mm,height = 80mm]{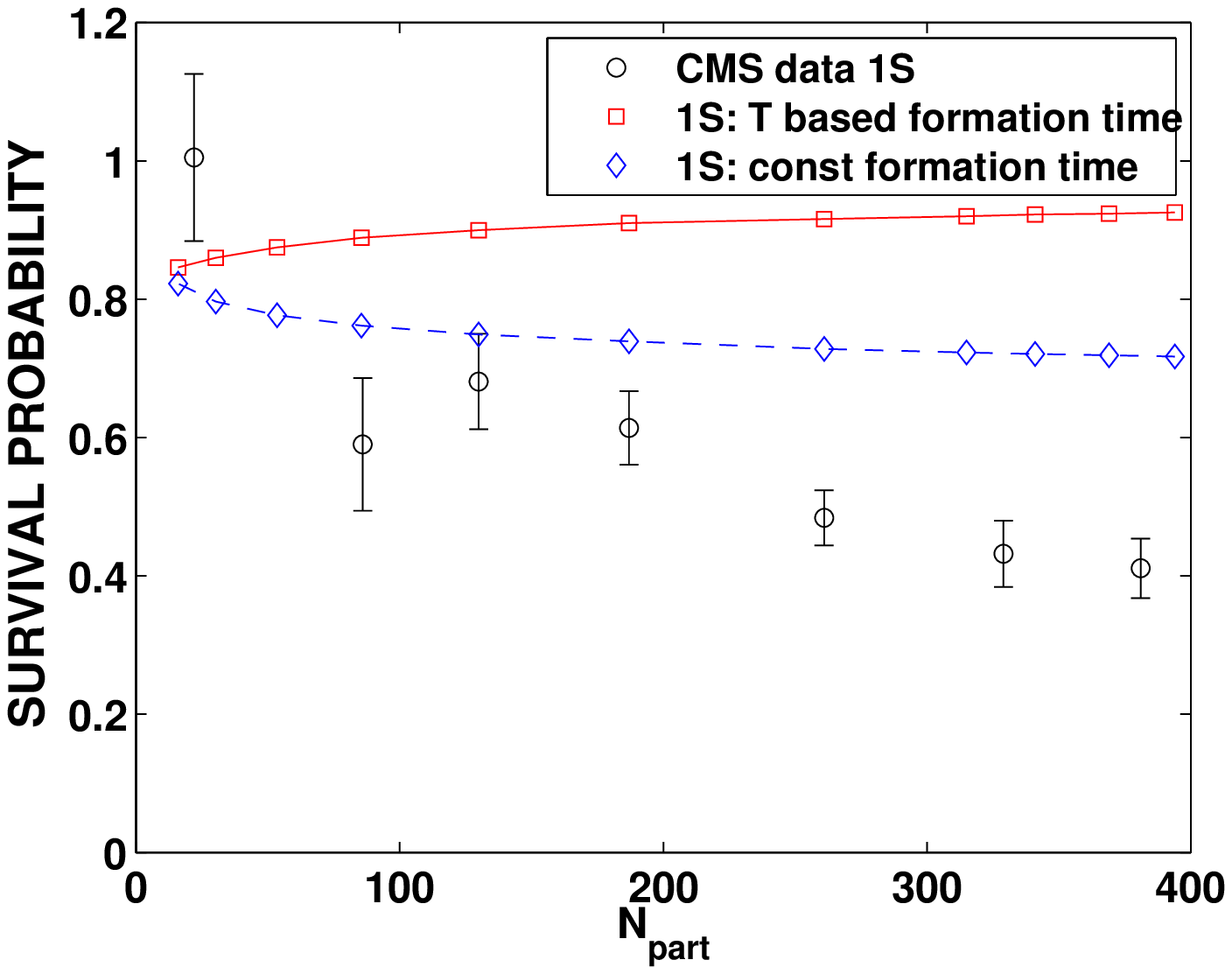}
\caption{CMS data~\cite{CMS2} compared with simulation results with only color screening for $\Upsilon(1S)$.}
\label{fig:1S_2S}
\end{figure}
\begin{figure}[h!]
\includegraphics[width = 80mm,height = 80mm]{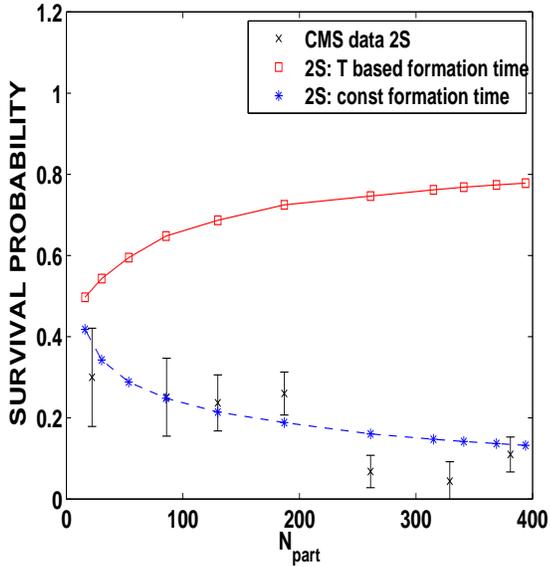}
\caption{CMS data~\cite{CMS2} compared with simulation results with only color screening for $\Upsilon(2S)$.}
\label{fig:1S_2S}
\end{figure}

\section{Results and Discussions}
The survival probability of the $\Upsilon(1S)$ and $\Upsilon(2S)$ states due to color screening versus the number of participants are depicted in Figs. 7 and 8, respectively. Each centrality region has different temperature and the formation time of both the $\Upsilon$ states has been modified based on the temperature corresponding to those centrality bins. The experimental CMS data in terms of the nuclear modification factor $R_{AA}$, measured as a function of the collision centrality~\cite{CMS2} are also shown in both the plots for comparison. 
The solid lines indicate the suppression for $\Upsilon(1S)$ and $\Upsilon(2S)$ in Figs. 7 and 8, respectively, due to color screening with temperature-dependent formation time. For comparison, the dashed lines in both figures indicate the suppression for $\Upsilon(1S)$ and $\Upsilon(2S)$ bottomonium states with constant formation time. 
In the peripheral region, where the modification in formation time is small, the suppression due to color screening remains similar to the suppression due to color screening with constant formation time. However, for central bins, where the formation time modification is significant, the modification in suppression due to color screening is also significant.
Qualitatively, the suppression now decreases while traversing from the peripheral region to the central region. At a higher temperature, the $\Upsilon$-wave-function stability criterion $\frac{d<r(t)>}{dt}=0$ is satisfied at a much later point when the separation between $b$ and $\bar{b}$ is much larger. Thus, despite weaker bonding due to color screening and increased distance between the bottom quark-antiquark pair, the quarkonia need not dissociate as long as the $b$, $\bar{b}$ pair separation is less than the critical distance $<r>$, where $\frac{d<r(t)>}{dt}=0$. This leads to decreased suppression in central collisions, where due to higher temperatures, the distance $<r>$ at which $\frac{d<r(t)>}{dt}=0$ happens is larger.
It is clear from the above figures that the temperature-dependent formation time model modifies the suppression upto some extent for $\Upsilon(1S)$ and is very significant for $\Upsilon(2S)$ suppression. The modification of formation time, thus becomes more critical for $\Upsilon(2S)$. It is also apparent from the above plots that the color screening model alone is not able to explain the experimental data on bottomonium suppression for central collisions.

\begin{figure}[h!]
\includegraphics[width = 80mm,height = 80mm]{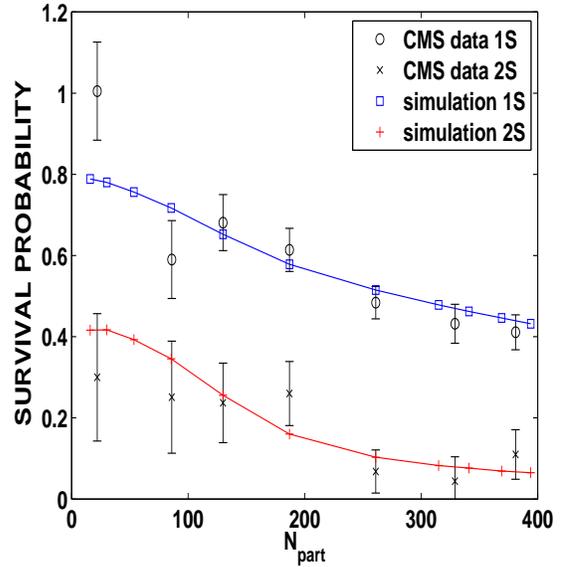}
\caption{Survival probability of $\Upsilon(1S)$ and $\Upsilon(2S)$ versus $N_{part}$ by incorporating color screening and gluonic dissociation with collisional damping. CNM effects are not incorporated. Temperature-dependent formation time using our proposed method is incorporated while calculating color screening.}
\label{fig:1S_2S_gdiss}
\end{figure}
In Fig. 9, we introduce gluonic dissociation and collisional damping in addition to the color screening mechanism. 
The difference between the CMS data and our model prediction has clearly decreased in comparison to Figs. 7 and 8. Our suppression results clearly overlap with the measured suppression data within error bars.
\begin{figure}[h!]
\includegraphics[width = 80mm,height = 80mm]{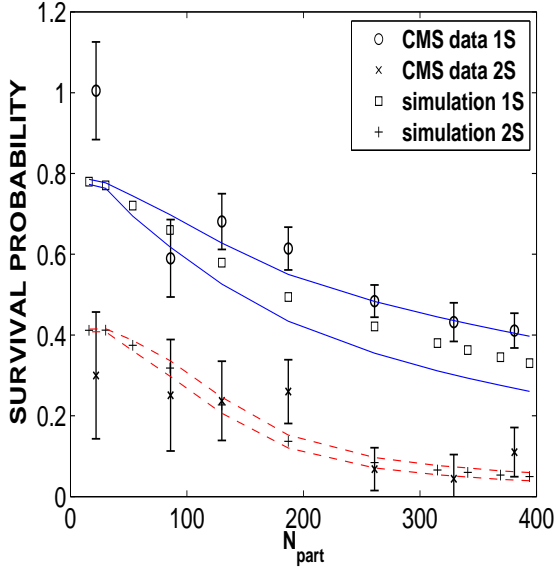}
\caption{Survival probability of $\Upsilon(1S)$ and $\Upsilon(2S)$ versus $N_{part}$ by incorporating color screening, gluonic dissociation and collisional damping along with CNM effects. The CNM calculation used here is based on the Vogt approach. The bands on the top and bottom indicate the maximum and minimum range of CNM for $\Upsilon(1S)$ and $\Upsilon(2S)$, respectively. The temperature-dependent formation time using our proposed method is incorporated while calculating color screening.}
\label{fig:1S_2S_all}
\end{figure}

In Fig. 10, the bottomonium suppression with all the above three effects, including CNM, is shown. This uses the first CNM approach, i.e., the Vogt approach. Our simulation results are in good agreement with the experimental data. The maximum and minimum range of CNM effects for $\Upsilon(1S)$ and $\Upsilon(2S)$ are indicated by the shaded region. The calculations are done using {\it EPS09} at leading order and with the temperature dependent formation time calculated by solving the time-independent Schr\"{o}dinger wave equation.
Comparison of the central region for $\Upsilon(1S)$ between Figs. 9 and 10 indicates that the mean of the experimental data falls between the suppression without CNM effects (shown in Fig. 9), and the mean suppression with CNM effects (shown in Fig. 10). The band depicting the possible CNM uncertainty overlaps with the error bars. This may indicate that the CNM may be on the lesser side of the CNM uncertainty band. The CNM uncertainty band for $\Upsilon(2S)$ is pretty small, but still a lesser value of the CNM effect continues to be in reasonable agreement with the experimental data.
  In Fig. 11, the variation of bottomonium suppression with respect to centrality using the temperature-dependent formation time obtained by simulation of the time dependent Schr\"{o}dinger equation is shown. Comparison of Fig. 10 with Fig. 11 shows that two different approaches to determine the temperature-dependent formation time give similar results. The differences in the peripheral region for $\Upsilon(2S)$ arises due to the differences in formation time in the peripheral region between the two methods as shown in Figs. 4 and 5. 
\begin{figure}[h!]
\includegraphics[width = 80mm,height = 80mm]{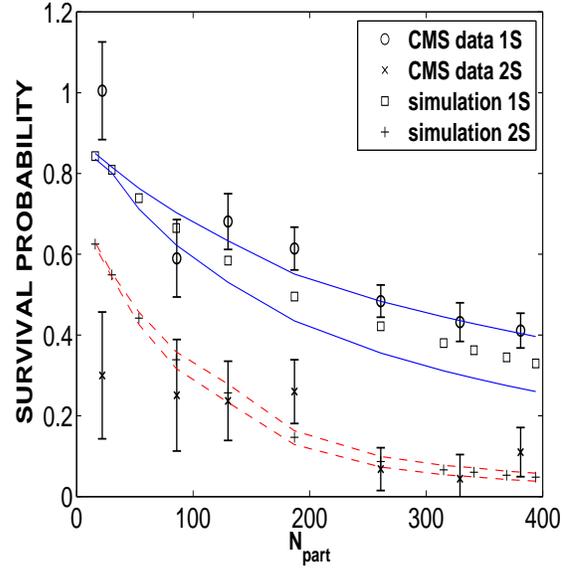}
\caption{Survival probability of $\Upsilon(1S)$ and $\Upsilon(2S)$ versus $N_{part}$ by incorporating color screening, gluonic dissociation and collisional damping along with CNM effects. The formation time  has been determined by simulation of the time-dependent Schr\"{o}dinger equation.}  
\label{fig:1S_2S_all}
\end{figure}

Figure 12 shows the comparison between the two CNM approaches for the variation of the final suppression of $\Upsilon(1S)$ and $\Upsilon(2S)$ states with $N_{part}$. 
\begin{figure}[h!]
\includegraphics[width = 80mm,height = 80mm]{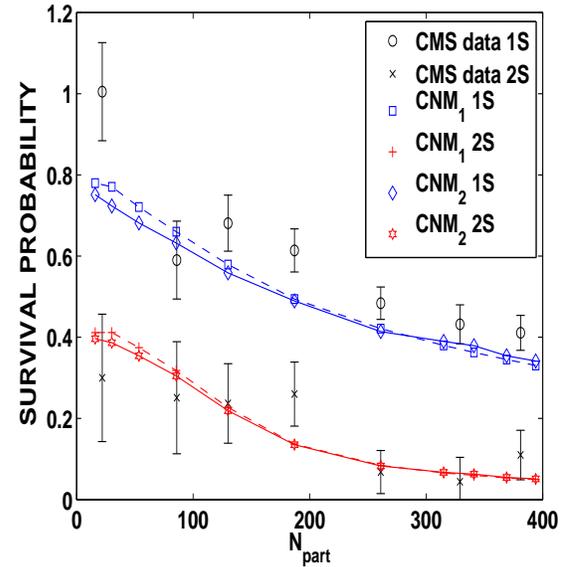}
\caption{Comparison between the two CNM approaches for $\Upsilon(1S)$ and $\Upsilon(2S)$. The subscript $1$ and $2$ in $CNM_1$ and $CNM_2$ refer to the two CNM approaches. $CNM_1$ is the "Vogt approach", while $CNM_2$ is "our approach".}
\label{fig:new_cnm_1S_2S}
\end{figure}
One can see from the above plot that there is little difference in the final suppression employing the two CNM approaches.
Shadowing has been shown to vary as $A^{1/3}$~\cite{ivanshad}. Even though we use only the neighboring nucleons, this shadowing methodology seems to give results upto a reasonable approximation. 
Figure 13 depicts our predicted suppression for $\Upsilon(1P)$ versus $N_{part}$ including all the effects employed in Figs. 10 and 11. The CNM model used here is based on the Vogt approach~\cite{vogt}. 
\begin{figure}[h!]
\includegraphics[width = 80mm,height = 80mm]{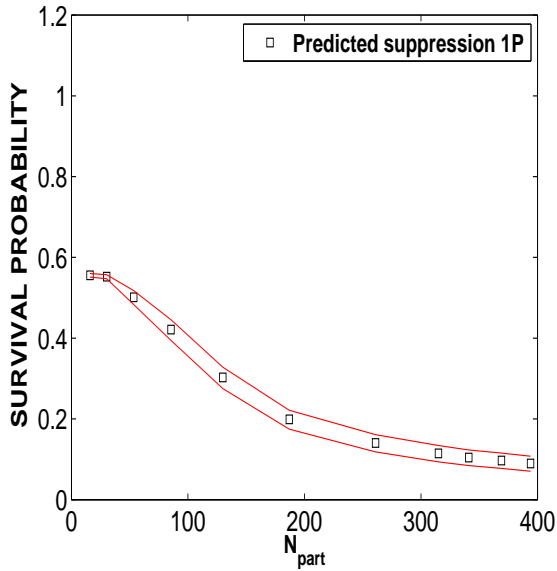}
\caption{Our prediction for the $\Upsilon(1P)$ bottomonium state using {\it EPS09} and {\it CTEQ6} after including color screening, gluon dissociation, collisional damping and the CNM effect.}
\label{fig:cnm_1P}
\end{figure}
Finally, Fig. 14 gives the comparison of the variation of overall $\Upsilon(1P)$ suppression with respect to centrality determined by employing two CNM approaches. Again, we see that both the CNM methods yield almost the same results. 
\begin{figure}[h!]
\includegraphics[width = 80mm,height = 80mm]{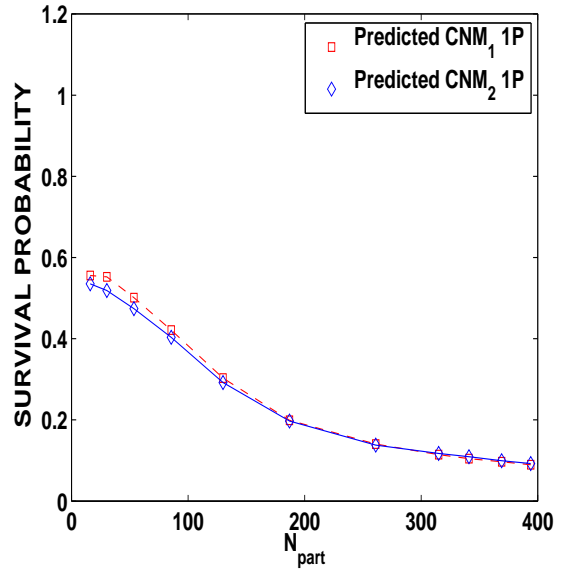}
\caption{Comparison between the two CNM approaches for $\Upsilon(1P)$. $CNM_1$ is the approach used by Vogt. $CNM_2$ is our approach.}
\label{fig:new_cnm_1P}
\end{figure}

\section{Conclusions}
  In conclusion, we have presented a comprehensive model of bottomonium suppression in the QGP medium by combining color screening, gluonic dissociation, and collisional damping. Temperature-dependent formation time of the bottomonium states is used in the current work. We have developed a method for estimating the modification of the formation time of the $\Upsilon$ with temperature by solving the time-independent Schr\"{o}dinger wave equation. Temperature dependent formation time has also been determined explicitly by simulating the temporal variation of the $\Upsilon$ wavefunction using the time-dependent Schr\"{o}dinger equation. We find that the two independent methods, based on different approaches, give comparable results.
The modification of formation time due to temperature modifies the $\Upsilon$ suppression to a considerable extent in the central region and plays a crucial role in accurately determining the $\Upsilon$ suppression.
The quasiparticle model is employed as an EOS for the QGP expanding under Bjorken's scaling law. 
The shadowing (as a CNM) effect has also been calculated by using two approaches, namely, the Vogt~\cite{vogt} approach and our approach. The final suppression of the bottomonium after taking into account the CNM effect is calculated as a function of the number of participants and the results are compared with the recent CMS data~\cite{CMS2} at the energies available at the LHC in the mid rapidity region. Our simulated results compare reasonably well with the CMS data. We also see that both the approaches for modeling the shadowing effect give similar results. 
 
\section*{ACKNOWLEDGMENTS}
One of the authors (S. G.) acknowledges Broadcom India Research Pvt. Ltd. for allowing the use of its computational resources required for this work. M. M. is grateful to the Department of Science and Technology (DST), New Delhi for financial assistance from the Fast Track Young Scientist project.

\end{document}